\documentclass[%
 reprint,
 amsmath,amssymb,
 aps,
prb,
]{revtex4-1}

\usepackage{graphicx}
\usepackage{dcolumn}
\usepackage{bm}
\usepackage{color}

\begin{document}


\title{Density of photonic states in aperiodic structures}

\author{Vladislav~A.~Chistyakov${}^{1}$}
\email{v.chistyakov@metalab.ifmo.ru}
\author{Mikhail~S.~Sidorenko${}^{1}$}
\author{Andrey~D.~Sayanskiy${}^{1}$}
\author{Mikhail~V.~Rybin${}^{1,2}$}

\affiliation{$^1$School of Physics and Engineering, ITMO University, St Petersburg 197101, Russia}
\affiliation{$^2$Ioffe Institute, St Petersburg 194021, Russia}

\date{\today}

\begin{abstract}
Periodicity is usually assumed
to be the necessary and sufficient condition for the formation of bandgaps, i.e. energy bands with a suppressed density of states.
Here, we check this premise by analyzing the bandgap properties of three 
structures that differ in the degree of periodicity
and ordering. We consider a photonic crystal, disordered lattice and ordered but non-periodic quasicrystalline structure. 
A real-space metric allows us to compare the degree of periodicity of these different 
structures.
Using this metric, we reveal that the
disordered lattice and the ordered quasi-crystal can be attributed to the same 
group of material structures. We examine the density of their photonic states both theoretically and experimentally.
The analysis reveals that despite their
dramatically different 
degrees of periodicity, the photonic crystal and the quasicrystalline structure demonstrate an almost similar suppression of the density of states. Our results give a new insight into
the physical mechanisms resulting in the formation of bandgaps.
\end{abstract}

\maketitle

\section{Introduction}
\label{sec:Intro1}

Modern condensed matter physics have stemmed from translational symmetry analysis, which determines the transport properties of waves according to the band structure, following the Bloch theorem, which is applicable to any periodic system. Thus, bandgaps related to the periodicity of dielectric index give photonic crystals a unique degree of control over electromagnetic waves. For example, photonic crystals enable localization of light inside small volume cavities, design of waveguides with specific dispersion properties, and modification of the light emission rate \cite{YablonovitchPRL1987,SajeevPRL1987,JoannopoulosNature1997,JoannopoulosBook2008}. These features make photonic crystals useful in optical applications; for instance, they are used as resonators \cite{YoungOE2014}, optical sensor \cite{Konorov:05}, and topological insulators for photons \cite{KhanikaevNatPhot2019}. However, certain restrictions on the choice of material and symmetry for photonic crystals limit the freedom in their design.  In particular, an omnidirectional band gap opens up only for a sufficiently high contrast of the refractive index and high symmetry of a structure. Reverse engineering method with a numerical optimization can be used to overcome this problem, but it often yields overly complex structures
\cite{HuangPRR2020,WernerOME2019,HuangATS2019,RodriguezNP2018}.

Another strategy to obtain bandgaps is to consider non-periodic systems. For example, disordered hyperuniform structures have the demanded bandgap
properties despite their random structure \cite{article,Man:13,PhysRevB.95.054119,PhysRevLett.127.037401}. Quasicrystalline structures are remarkable among other non-periodic systems, because, despite the lack of translational symmetry, they are ordered. Recent advances in 3D nanomanufacturing enabled the fabrication of these structures with a fine precision allowing them to 
operate even in the visible light range~\cite{xu2007icosahedral,SprafkeAOM2018,EichOE2018,PetrovAPLP2020,NorrisNat2020,RybinAOM2020,koreshin2022interlaced}. 

Quasiperiodic photonic structures are a special class of photonic structures with a high degree of order in reciprocal space. The opportunity to design structures in the reciprocal space and then obtain their structure in real space by inverse Fourier transform attracts considerable attention \cite{NorrisNat2020}. Real-space methods allow engineering photonic structures with the desired properties, for example, complex moir\'{e} patterns \cite{wang2020localization}. Recently, quasicrystalline structures based on multiple one-dimensional (1D) gratings oriented in all directions were reported \cite{PetrovAOM2022}. Even a weak dielectric contrast in case of polymer materials allows such structures to exhibit an unprecedented suppression of radiation due to the bandgaps created in 2D and 3D cases.

\begin{figure}
\includegraphics{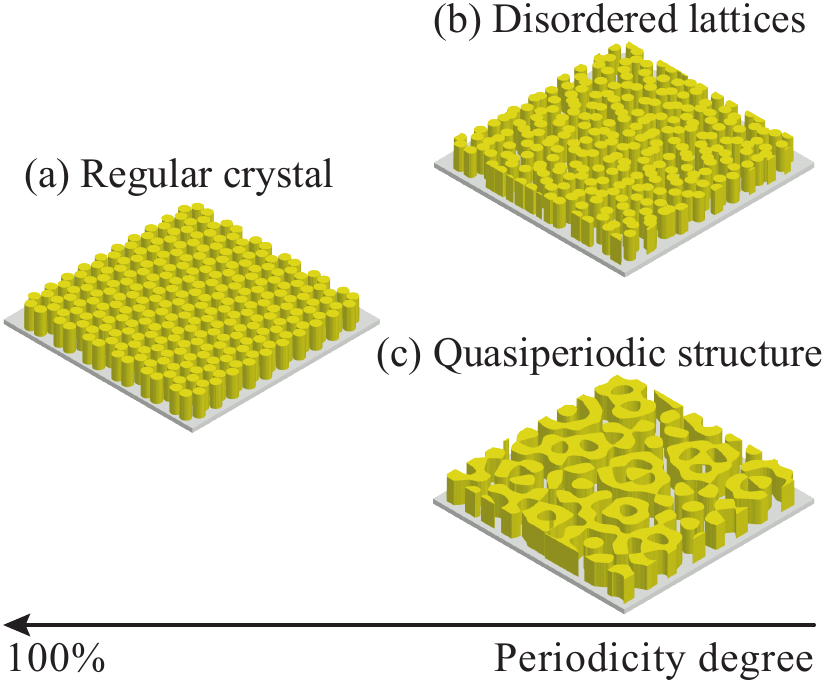}
\caption{\label{fig:ordVsperiod}
Landscape of ordered and disordered structures with different degrees of periodicity. 
(a) Honeycomb photonic crystal, an ordered periodic structure. (b) Disordered lattice with 
rods randomly shifted with respect to the rod positions in 
the honeycomb lattice, a disordered and non-periodic structure. (c) Quasiperiodic structure, 
an ordered but non-periodic structure.}
\end{figure}

Translational symmetry allows classification of solutions by their wave vector: such solutions are either a propagating or evanescent wave \cite{JoannopoulosBook2008}. In the latter case, the wave vector has a non-zero imaginary part and for a certain direction, there is a band gap. 
In disordered lattices, there is another source of evanescent solutions related to the Anderson localization~\cite{AndersonPhysRev1958,LopezAOM2018}. Figure~\ref{fig:ordVsperiod} shows an ordered photonic crystal with a periodic structure, a non-periodic ordered quasicrystal, 
and a disordered lattice. The quasicrystalline structure lacks both conditions for evanescent waves, because it has neither periodicity nor disorder.

In this paper, we study the suppression of light in a quasicrystalline structure and compare it with similar properties obtained in a photonic crystal with a honeycomb lattice, and in a disordered lattice structure. By using a real-space metric, we determine the amount of periodicity that allows us to numerically compare the quasiperiodic structure with the disordered lattice with a given degree of disorder. Numerical results reveal a strong suppression of wave propagation in the quasicrystalline structure comparable to its photonic crystal counterpart; in contrast, for the disordered lattice with the same degree of periodicity, the suppression is weaker. The predicted properties are farther confirmed by microwave transmission measurements.

\section{\label{sec:chapter1} Design of photonic structures}

Here, we describe photonic structures with a spatial distribution of permittivity modulated in two dimensions. We start with rigorously ordered quasiperiodic structures with no translational symmetry, then we study a photonic crystal with perfect ordering and periodicity, and then we consider a disordered non-periodic lattice.

Following Ref.~\onlinecite{PetrovAOM2022}, we generate  the quasiperiodic structure as a superposition of several one-dimensional gratings with a sine-type modulated refractive index. The grating orientations are uniformly distributed over the polar angle, and
the continuous spatial distribution of the refractive index is:
\begin{equation}\label{eq:structure}
\Delta n_{c} ({\mathbf r}) = \sum_{i=1}^{N_{opt}}\Delta n_i\,\mathrm{sin}\left({\bf b}_i\cdot {\bf r} + \phi_i \right),
\end{equation}
where $i$ enumerates the gratings, $N_{opt}$ is the optimal number of gratings, $\Delta n_i$ is the modulation amplitude of the refractive index in the $i$-th grating, ${\mathbf b}_i$ is the vector determining the grating period and orientation, and $\phi_i$ is a random phase. For such quasicrystalline structures, the optimal number of gratings is $
    N_{opt} \approx 2.36\left( 
    \overline{n}/\Delta n\right)^{2/3}$,
where $\overline{n}$ is the mean refractive index and $\Delta n$ is the amplitude of the refractive index deviation from the average value $\overline{n}$, i.e. $n_1 = \overline{n} +\Delta n$ and $n_2= \overline{n}-\Delta n$.

\begin{figure}
\includegraphics{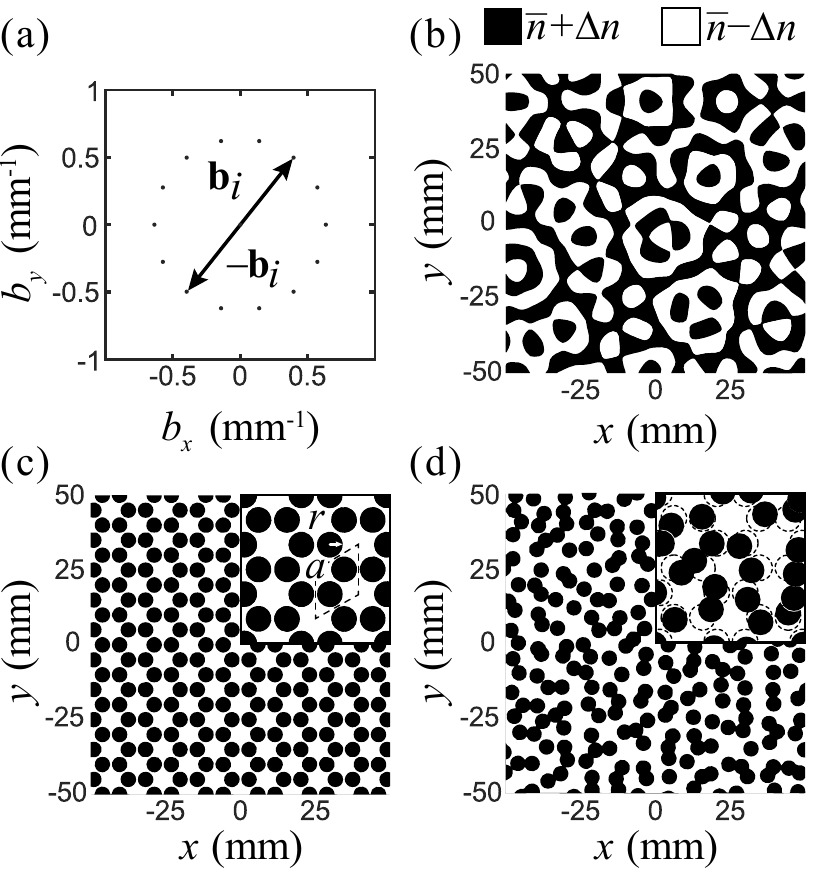}
\caption{\label{fig:reciprocal}
(a) Schematics of fourteen Bragg maxima located along the circle in the reciprocal space corresponding to the seven gratings merged into a single structure, which is
used for generating a quasiperiodic structure with an omnidirectional Bragg bandgap. (b) Fragment of the quasiperiodic structure with a binary distribution of two materials. The structure corresponds to seven gratings with the distribution of the maxima in reciprocal space shown in panel (a). (c) Ordered honeycomb photonic crystal composed of dielectric rods in air. (d) Disordered lattice of dielectric rods in air. The rod positions are randomly shifted in both $x$ and $y$ directions with respect to the nodes of the honeycomb lattice.
}
\end{figure}

Optical properties of the structure can be analyzed in the reciprocal space strictly connected to Fourier transform. The Fourier transform of the sine function corresponds to two Bragg maxima (Dirac delta functions) with the direction $\pm {\mathbf b}_i$ in reciprocal space. Each lattice has the same period $a$, so the lattice vector length is defined by $b = |{\mathbf b}_i | = 2\pi /a$. The uniform angular distribution of the gratings leads to the superposition in (\ref{eq:structure}), which results in dense distribution of first-order Bragg maxima around a circle in reciprocal space, shown in Fig.~\ref{fig:reciprocal}a. Because of randomly chosen phases and sufficient number of gratings, the local perturbation of the refractive index has a homogeneous distribution with no remarkable features in any specific direction (see Fig.~\ref{fig:reciprocal}b). Notably, according to the recent report~\cite{PetrovAOM2022}, such a quasiperiodic structure has a complete photonic band gap for all directions of propagation of the electromagnetic wave.

To obtain a structure that comprises only two dielectric materials, a binarization procedure \cite{BerkPhysRevA1991,LinOE2005,SteinhardtPhysRevLett2008,10.1117/12.662380,IvanovOE2004} is applied to the refractive index sum $\Delta n_{c} ({\bf r})$
\begin{equation}
    n_b({\bf r}) = \overline{n} + \Delta n \mathrm{sign}\left(\Delta n_{c}({\bf r})  \right).
\end{equation}
After binarization, each grating contributes to the light scattering with an effective amplitude $\Delta n_{b,i} = 2\Delta n / \sqrt{\pi N_{opt}}$. The binarization procedure introduces additional noise in the reciprocal space, and its integral background value is 
36\% 
of the Bragg maxima intensity. However, for a sufficiently high grating number, these maxima are so densely distributed that they are almost indistinguishable in a finite-size structure~\cite{EichOE2018_2} (see Fig. \ref{fig:reciprocal}a). Therefore, a proper choice of the grating number provides a sufficient density of the maxima along the circle. 

We chose a photonic crystal with a honeycomb lattice as a periodic and ordered structure (Fig.~\ref{fig:reciprocal}c). Each hexagonal unit cell contains six dielectric rods with a radius of $r=0.26 a$ surrounded by air. The filling fraction of the rod material matches that of the quasiperiodic structure. Both structures have the same lattice constant $a$. The disordered lattices with no translational symmetry \cite{DrouardOE2016,HughesPRA2015,NodaAPL2012,HickmannJAP2008} are based on the honeycomb photonic crystal, but the position of each rod was randomly shifted according to $x=x_{0,j}+\delta x_j$ and $y=y_{0,j}+\delta y_j$. Here, $\delta x_j$ and $\delta y_j$ are random values distributed uniformly in the interval $[0\dots d_r]$. Fig.~\ref{fig:reciprocal}d displays an example of such a disordered lattice.

\section{Metric for periodicity degree}
\label{sec:chapter3}

Since we aim to reveal how periodicity affects the structures' properties, a proper metric is required for quantitative analysis. Such a metric would allow us to compare ordered quasiperiodic structures with photonic crystals with a certain value of disorder 
$\sigma$. Currently, there exists no standard metric with general qualitative and quantitative properties for an arbitrary case. Various metrics were developed and applied to a wide range of disordered systems \cite{ParkNature2020}. Most of the proposed metrics use the characteristics of the disorder spectrum in reciprocal space \cite{KlattPhysRevX2021,TorquatoPhysRevE2020,StillingerPhysRevE2003,StillingerPhysRevX2015}. However, they are inefficient for the description of samples that are small in real space. Although several studies proposed non-trivial real-space metrics that allow distinguishing structures in terms of pseudodisorder \cite{DrouardOE2016,HughesPRA2015}, they cannot be generalized for other classes of photonic crystals, including quasiperiodic structures. Thus, an ad hoc approach is required.

\begin{figure}
\includegraphics{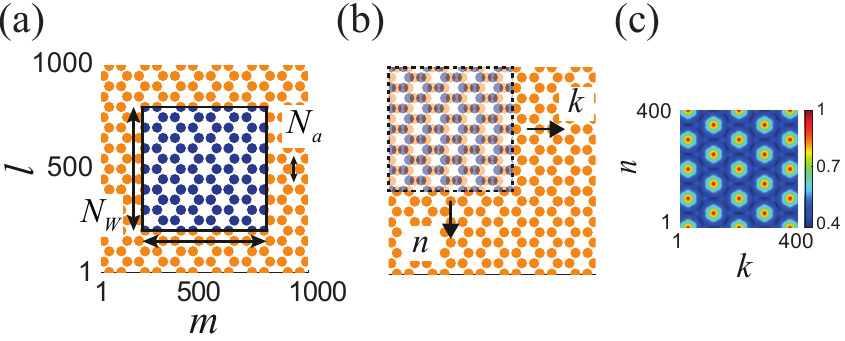}
\caption{\label{fig:convolut}
Illustration of calculation of a discrete convolution for the ordered photonic crystal. (a) Choosing a finite area $W$ in a complete structure $F$. (b) Calculation of the discrete convolution values. (c) Convolution of the ordered photonic crystal.
}
\end{figure}

In this study, we exploit a metric based on self-convolution in real space. A reliable figure of merit is obtained by numerical convolution of the sample with itself \cite{ZouRemSens2020}, which means calculating the structure numerically after discretizing the real-space domain into finite square samples
\begin{equation}
    C_{2D}(n, k) = \sum_{i=1}^{N_W} \sum_{j=1}^{N_W} W(i,j)  F(n+i-1, k+j-1),
\end{equation}
where $n = 1\dots N_F-N_W+1$, $k = 1\dots N_F-N_W+1$ are the elements of the output convolution matrix, $F(l,m)$ is an
$N_F$ by $N_F$ binary matrix describing the complete structure, $W(i,j)$ is an $N_W$ by $N_W$ area
of the complete matrix of the structure for $(N_F/N_a-4)$ periods with the lattice constant 
$N_a = a/\Delta$, and $\Delta = 0.1$ is the sampling step. Fig.~\ref{fig:convolut} shows such a 2D convolution for the ordered photonic crystal. At each location, we calculate the product of
each element of the square area and each element of the complete matrix, i.e. we find their overlap, and then we sum the results to obtain the output in the current location \cite{dumoulin2016guide}. Figs.~\ref{fig:mapconvolut}a and~\ref{fig:mapconvolut}b show the convolution maps for the considered disordered lattice and quasiperiodic structure, respectively. The maxima in the maps reflect the 
degree of periodicity of a structure with a specific lattice constant $a$. To quantify the periodicity degree, we consider the nearest maxima located at a distance from the center equal to the lattice constant $a$. We notice here that for the quasiperiodic sample, the maxima merge and form a ring, which indicates an omnidirectional band gap. The non-periodic distribution of structures is manifested in the broadening of the maxima on the map along with the decrease in their intensity.

\begin{figure*}
\includegraphics{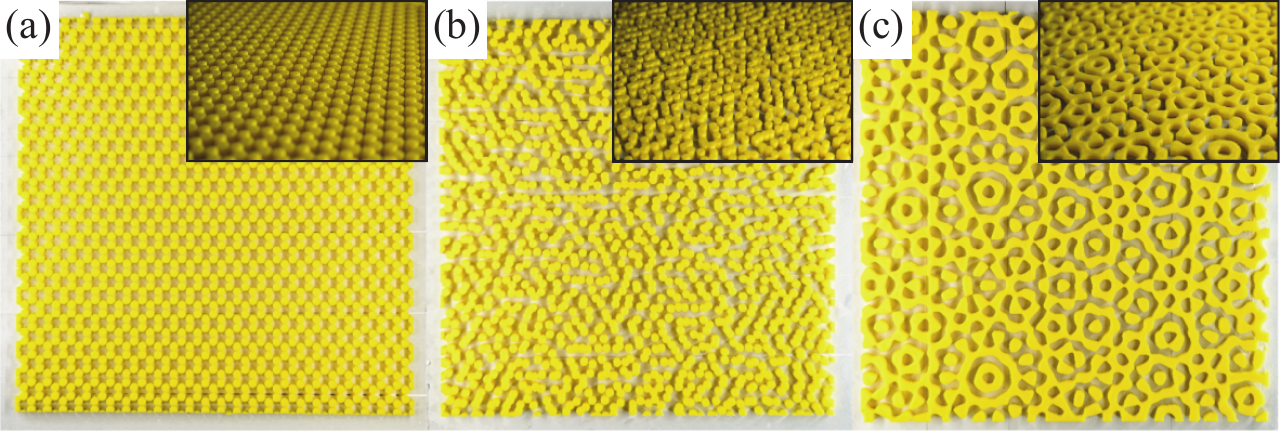}
\caption{\label{fig:samples}
Photographs of samples printed on a 3D printer. (a) Ordered honeycomb photonic crystal of dielectric rods in air. (b) Disordered lattice of dielectric rods in air. The rod positions are randomly shifted away from the honeycomb lattice positions in both $x$ and $y$ directions (the disorder degree is $d_r=2$ mm). (c) Quasiperiodic structure with a binary distribution of two materials. The structure corresponds to seven gratings merged into a single binary structure.
All the structures are 5 mm high. Structures in (b) and (c) have the same degree of periodicity  $\sigma=4.78$.
}
\end{figure*}

At the next step, we consider a one-dimensional normalized convolution $C_\mathrm{1D} (x)$ for the selected maxima of both structures, shown in Fig.~\ref{fig:mapconvolut}c with solid curves. These data are described adequately by using a second convolution of Gaussian $G(x)$ and rectangular $\mathrm{rect}(x)$ functions, and the disorder is quantified with respect to the standard deviation of the Gaussian function. The corresponding equation read
\begin{eqnarray}
    C_{rG}(x) = \sum_{x=1} \mathrm{rect}(x)  G(k-x+1), \\
    C_\mathrm{1D}'(x) = \sum_{x=1} C_{rG}(x)  C_{rG}(k-x+1) \approx C_\mathrm{1D}(x), 
\end{eqnarray}
where $\mathrm{rect}(x)$ is a unit-height rectangular pulse of length $C_\mathrm{1D} (x)$, $G(x)=\exp(-x^2/(2\sigma^2))/(\sigma\sqrt{2\pi} )$, and $\sigma$ is the standard deviation. By applying the one-dimensional convolution to our structures, we find the standard deviation $\sigma$ for the Gaussian distribution when $C_\mathrm{1D}'(x) \approx C_\mathrm{1D}(x)$. The larger the structure and the square area are, the more stable the standard deviation value is. For our ordered
photonic crystal, the dispersion $\sigma = 0$, which is proven by the fact that the corresponding convolution function is triangular, see the green curve in Fig.~\ref{fig:mapconvolut}c. The dashed curve in Fig.~\ref{fig:mapconvolut}c shows the metric function $C_\mathrm{1D}' (x)$ for the disordered lattice. This function describes the original maximum in the map, therefore, we can apply this metric to describe the degree of periodicity 
in random photonic crystals as well.

We find the minimum of the RMSE function~\cite{DraxlerGeo2014} of the difference in the convolution maps of the quasiperiodic structure and disordered lattice. By this criterion, the quasiperiodic structure equivalent to the disordered lattice with a degree of disorder of $d_r=2$ mm, as can be seen in the inset in Fig.~\ref{fig:mapconvolut}c. In this case, the described metric reveals a standard deviation of $\sigma=4.78$ for both structures. Thus, these two structures have the same degree of periodicity, which is the parameter responsible for the formation of photonic band gaps.

\begin{figure*}
\includegraphics{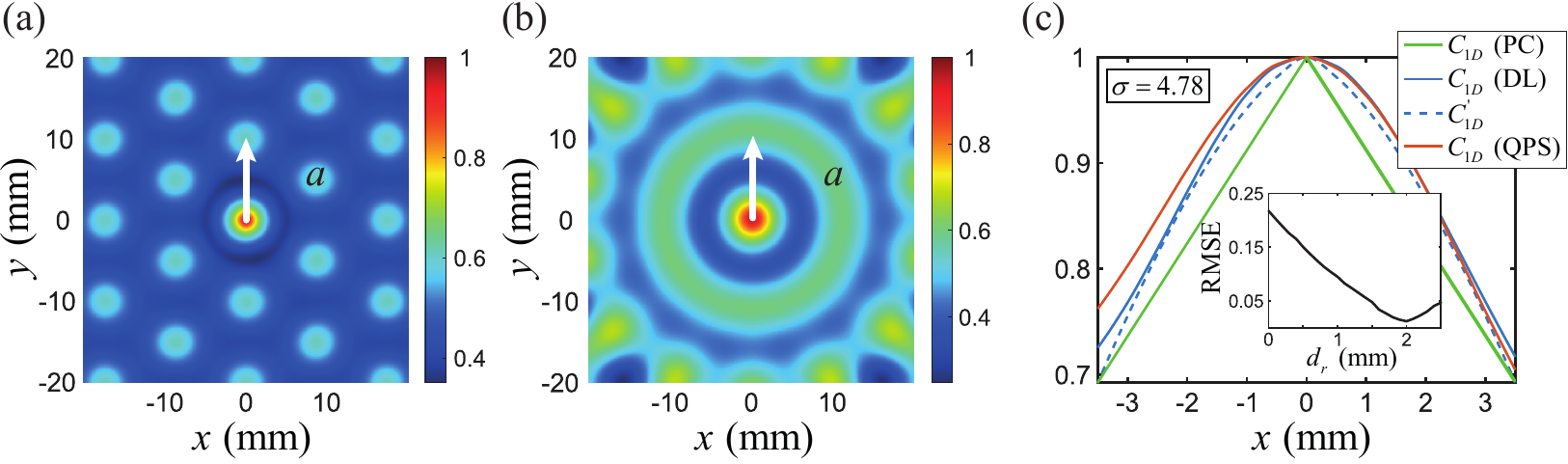}
\caption{\label{fig:mapconvolut}
Convolution maps of the disordered lattice (a) and the quasiperiodic structure (b) shown in Figs.~\ref{fig:samples}b and ~\ref{fig:samples}c, respectively. The vectors of translation are shown with white arrows.
(c) Comparison of the cut of convolution along the $x$-axis for the honeycomb photonic crystal 
(green curves), the quasiperiodic structure (red curve), the disordered lattice
(blue curve) and the metric function for the disordered lattice (blue dashed curve). Inset: 
RMSE difference between the convolutions of the quasiperiodic structure and the disordered lattice as a function of the degree of disorder of the disordered lattice.}
\end{figure*}

\section{Probing of Band gap properties}
\label{sec:chapter4}

\begin{figure*}
\includegraphics{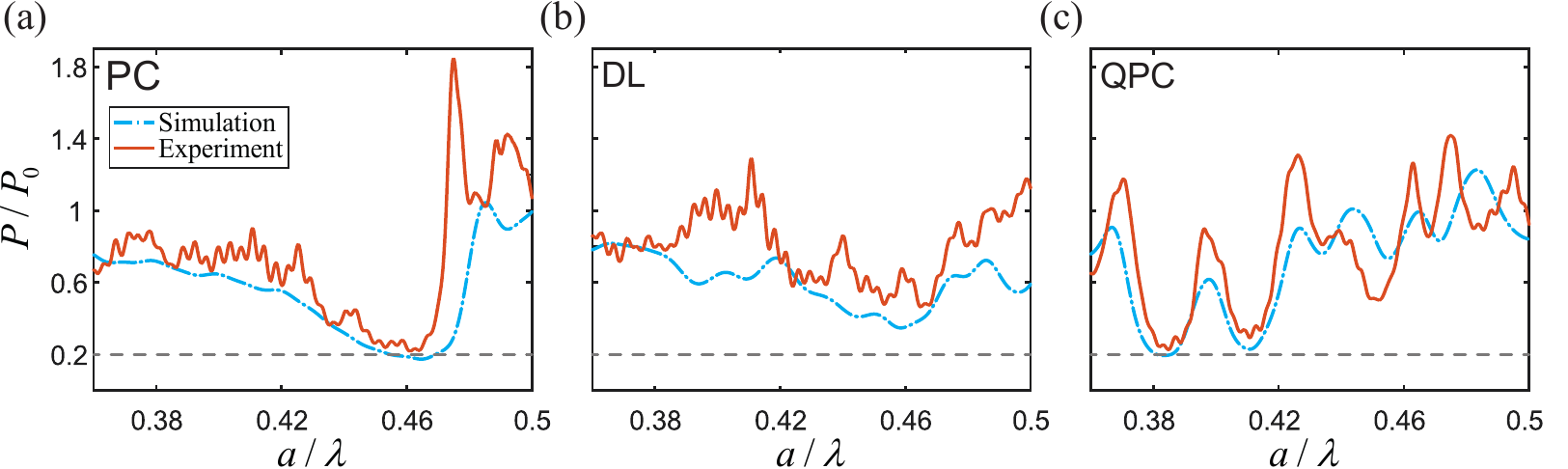}
\caption{\label{fig:results}
Normalized radiation power $P/P_0$ of a dipole placed inside the 
honeycomb photonic crystal (a), the disordered lattice (b) and the quasiperiodic structure (c). The plots correspond to the theory (dashed blue curve) and the experiment (solid red curve). The gray dashed line shows the level of radiation suppression up to 0.2.
}
\end{figure*}

Now we compare the electromagnetic properties of the quasiperiodic structure and the disordered lattice with the same degree of periodicity according to the metric described above. First, we carried out full-wave simulation by using the time-domain solver of CST Studio Suite software. A linear dipole emitter was located at the center of each structure under consideration. We also analyzed the results after averaging the data for 15 different dipole positions in the central unit cell. To facilitate further experiments, we chose a dipole source with TM polarization (in this case, electric field oscillates along the vertical $z$ direction). The top and bottom boundaries of the structure were chosen to be a perfect electric conductor; the vertical boundaries of the structure in the $x$-$y$ plane were a perfectly matched layer to simulate open boundary conditions.

To verify our theoretical predictions in experiment, we used polymer samples fabricated with
a 3D printer. A pair of aluminum plates formed a plane-parallel waveguide for TM polarization, and the samples were placed between the plates. A small dipole antenna was located in the center of the structures. We measured the parameters of the structures in the frequency range from 10.8 to 15 GHz. For each structure, the reflection coefficient ($S_{11}$ parameter) was measured using a vector network analyzer, and the real part of $S_{11}$ allowed estimating the emitted power $P$. To measure the reference radiation power in vacuum $P_0$, we placed the dipole antenna in the plane-parallel waveguide without the structure.

For the structures both in the simulation and in the experiment, we chose polylactide (PLA) plastic with a permittivity of about $\varepsilon_1 = 2.2$ ($n_1 \approx 1.5$) and low losses ($\tan \delta \approx 1\cdot10^{-2}$) in the microwave range \cite{LimMaterials2019}. The dielectric structures were surrounded by air $(n_2 \approx 1)$. The optimal number of gratings for the quasiperiodic structure with the refractive contrast $n_1/n_2 =1.5$ was $N_{opt} = 7$. The scale of the structures was chosen for the lattice constant to have the same value, $a=10$ mm, at different frequencies of the band gap of the quasiperiodic structure and the photonic crystals. For the photonic crystals, the radius of the dielectric rods was $r=0.26 a$.  These structures are shown in Fig.~\ref{fig:samples}. The structures had the dimensions of $250 \times 250 \times5$~mm$^3$ ($25a \times 25a \times 0.5a$), and the total volume filling fraction was
about 50\% for all the samples.

First, we simulated the radiation efficiency and studied the radiation suppression due to the modification of the local density of photonic states. The emitted power of the dipole $P$ was estimated by real part of the impedance normalized to that of the dipole located in free space $P_0$~\cite{ BelovSciRep2015,MendachPhysRevLett2016, EichOE2018}. For convenience, we use dimensionless frequency $a/\lambda$ for representation of the emission spectra of the dipole in the structures under study (the dashed blue curve in Fig.~\ref{fig:results}). The photonic crystal structures exhibit a suppression band around $0.46a/\lambda$ (Fig.~\ref{fig:results}a). A pair of emission suppression gaps is clearly seen around $0.38a/\lambda$ and $0.41a/\lambda$ in the power spectrum of the quasiperiodic structure (Fig.~\ref{fig:results}c). We notice that the spectrum of disordered lattice is quite similar for different realizations of the structure. Thus, we do not average the spectrum over the ensemble. The suppression of the radiation related to the local density of photonic states decreases strongly within the band gaps. Noteworthy, the finite sizes of the structures result in non-zero local density of photonic states in the band gap. Macroscopic interference effects cause oscillation fringes that are observed in all the spectra at the frequencies outside the bandgaps, but these features are neglected in the further discussion. An important observation is that the both regular structures demonstrate almost the same suppression of radiation down to the baselevel of 20\% despite the rather distinct degree of periodicity. The disordered lattice with the same degree of periodicity as the quasiperiodic structure demonstrates a significantly weaker suppression, about 40\% (Fig.~\ref{fig:results}b).

The normalized TM power emission spectra of the dipole obtained in experiment for each structure are shown by the solid red curve in Fig.~\ref{fig:results}. In contrast to the simulation, high-frequency fluctuation fringes are observed in the spectra for all the structures. These fringes 
are likely to result from the spatial gaps between the sample and metal planes in the experimental setup. These gaps create additional waveguide modes, which are absent in the simulations. However, such oscillations do not affect the analysis of the bandgap suppression features.

For all the three structures, the spectra exhibit bandgaps around $0.46a/\lambda$ (photonic crystal samples in Fig.~\ref{fig:results}a) and $0.38a/\lambda$ and $0.41a/\lambda$ (the quasiperiodic sample in Fig.~\ref{fig:results}c), which is in excellent agreement with the ones obtained in simulations. Moreover, the suppression of the electromagnetic radiation due to the Bragg band gaps is stronger for the ordered structures (down to the baseline at 20\%). In the experiment, the disordered lattice (Fig.~\ref{fig:results}b) with the same degree of periodicity as that of the quasiperiodic structure demonstrates weaker effects related to the emission suppression (with a baseline at about 40\%). Thus, the experimental data completely confirms the theoretical predictions.

Let us discuss the parameters related to the modification of the local density of photonic states. The main parameters contributing to this effect are the following: the size of the structure, dielectric contrast, stop bands overlapping for all spatial directions, and the degree of periodicity. The translational symmetry of a crystal lattice limits possible rotational symmetries, which are responsible for the overlap of the stop
bands in all directions. If the dielectric contrast is low, there is an additional leakage of radiation in photonic crystals, and non-periodicity introduces another leakage channel. In contrast, quasiperiodic structures have no strict restrictions on the overlapping of stop bands, but a fixed degree of periodicity opposes the suppression of the local density of states. These competing effects create opportunities for designing polymer-based structures with a local density of states due to an additional degree of freedom provided by rotational symmetry. 

\section{Conclusion}
\label{sec:Conclusion}

In this work, we have analyzed the opportunities provided by quasiperiodic structures for advanced manipulation of electromagnetic radiation. Our study focuses on the suppression of local density of photonic states and its interplay with the degree of periodicity of the structures. We have proposed a real-space metric to compare the photonic properties of quasiperiodic structures 
and disordered lattices. We have found that quasiperiodic structures made of available plastic materials achieve the same suppression of local density of photonic states as photonic crystal with a perfect translational symmetry does. We have carried out both theoretical and experimental research, and the results are in excellent agreement. Surprisingly, we have revealed that the lack of periodicity is beneficial for the suppression of density of photonic states, and non-periodic structures have inherent advantages over the ordered ones in this regard. 
Our findings pave the way for engineering photonic structures made of various low-index materials with an additional degree of freedom enabled by quasicrystal design. We anticipate that polymer-based structures empowered with a quasiperiodic topology will broaden the possible  applications of photonic structures.

\begin{acknowledgments}
We thank Alexander Petrov for fruitful discussions.
Authors acknowledge a support from the Russian Science Foundation (Grant No. 20-79-10316).   
\end{acknowledgments}

\bibliography{QuasiDisorder}

\end{document}